\documentclass[aps,prl,twocolumn,amsmath,amssymb,nofootinbib,floatfix,superscriptaddress]{revtex4-1}

\usepackage[T1]{fontenc}
\usepackage{textcomp}
\usepackage{amsmath,braket}
\usepackage{amssymb}
\usepackage{amsthm}
\usepackage[dvipdf]{color}
\usepackage{graphicx}
\usepackage{dcolumn}
\usepackage{float}
\usepackage{bm,subfigure} 
\usepackage{xcolor,colortbl}
\usepackage[normalem]{ulem}
\usepackage{lipsum}
\makeatletter
\renewcommand{\@makefntext}[1]{%
  \noindent\makebox[1.8em][l]{\@thefnmark}#1}
\makeatother

\DeclareMathOperator{\tr}{tr}

\newtheorem{theorem}{Theorem}[section]
\newtheorem{lemma}{Lemma}[section]

\newcommand{\RR}{\mathbb{R}}
\newcommand{\im}{\operatorname{im}}

\begin{document}

\title{Failure of the Goldstone Theorem for Vector Fields and Boundary-Mode Proliferation in Hyperbolic Lattices}

  \author{Daniel Sela$^*$}
  \author{Nan Cheng$^*$}
  \author{Kai Sun}
\affiliation{
 Department of Physics, University of Michigan, Ann Arbor, MI 48109-1040, USA
}

\begin{abstract} 
Hyperbolic lattices extend crystallinity into curved space, where negative curvature and exponentially large boundaries reshape collective excitations beyond Euclidean intuition. In this Letter, we push the study beyond scalar fields by exploring vector fields on hyperbolic lattices. Using phonons as an example, we show that the Goldstone theorem breaks down for vector fields in hyperbolic lattices. In stark contrast to Euclidean crystals, where the Goldstone theorem ensures that acoustic phonon modes are gapless, hyperbolic lattices with coordination number $z > 2d$ exhibit a finite bulk phonon gap. We identify the origin of this breakdown: the Goldstone modes here belong to nonunitary representations of the translation group and therefore cannot form gapless excitation branches. We further show that when boundaries are included, this bulk spetrum gap is filled by an extensive number of low-frequency boundary modes.
\end{abstract}
\maketitle
\def\thefootnote{*}
\footnotetext{These authors contributed equally to this work.}
\def\thefootnote{\arabic{footnote}} 

\noindent{\it Introduction}---Extending crystallinity into curved space, hyperbolic lattices break the geometric constraints of Euclidean crystals and unveil new collective and topological phenomena that defy flat-space intuition. Theoretically, these lattices have opened new avenues of exploration, spanning hyperbolic band theory~\cite{Maciejko_2021,Maciejko_2022,Cheng_2022,urwyler2022hyperbolic,lenggenhager2023non,shankar2024hyperbolic}, hyperbolic topological states~\cite{Liu_2022,zhang2022observation,zhang2023hyperbolic,chen2024anomalous,Guan_2025,He_2024}, hyperbolic spin liquids~\cite{Lenggenhager_2025,Dusel_2025}, non-Abelian semimetals~\cite{tummuru2024hyperbolic}, and Anderson localization on curved lattices~\cite{chen2024anderson}, revealing new classes of many-body systems shaped by geometry and curvature. Beyond condensed matter, hyperbolic lattices also serve as powerful frameworks for quantum error-correcting codes~\cite{pastawski2015holographic} and holographic simulations~\cite{dey2024simulating}. Experimentally, they have been realized across diverse platforms, including circuit quantum electrodynamics~\cite{Koll_r_2019}, classical electrical circuits~\cite{lenggenhager2022simulating,chen2023hyperbolic}, and photonic systems~\cite{huang2024hyperbolic1,huang2024hyperbolic2,park2024scalable}.

Unlike Euclidean Bravais lattices, hyperbolic lattices are characterized by two intertwined geometric features—negative curvature and a finite boundary-to-bulk ratio~\cite{iversen1992hyperbolic}—which together control their physical behavior.
Although existing studies have made significant progress in understanding scalar fields on hyperbolic lattices, particularly under periodic boundary conditions~\cite{Maciejko_2021,Maciejko_2022,chen2023hyperbolic,Guan_2025,He_2024,lenggenhager2022simulating,mosseri2023density,lux2023converging}, two key features that fundamentally distinguish hyperbolic from Euclidean lattices require further investigations—(1) what new physics emerges from vector fields, and (2) how extensive boundaries reshape the bulk–edge relationship.

The first concerns the influence of curvature on vector fields. Unlike scalar fields, which remain invariant under parallel transport, a vector field directly senses curvature through holonomy~\cite{do2016differential}: transporting a vector around a closed loop rotates its orientation, exposing the underlying geometry.
The second arises from the finite boundary-to-bulk ratio inherent to hyperbolic lattices~\cite{iversen1992hyperbolic}, in stark contrast to Euclidean crystals where this ratio vanishes in the thermodynamic limit. As a consequence, conventional notions such as bulk–edge correspondence must be revisited, as their foundational assumption of a negligible boundary no longer holds.

In this Letter, we investigate phonons—vector-field excitations—on hyperbolic lattices, focusing on both bulk and boundary modes. Obtaining the full phonon spectrum within conventional hyperbolic band-theory frameworks requires determining all high-dimensional irreducible representations of the hyperbolic translation group~\cite{Cheng_2022}, rendering analytical calculations intractable. To overcome this difficulty, we develop an alternative moment-based approach that extends earlier methods~\cite{mosseri2023density} for scalar models and remains accurate in the low-frequency regime where continued-fraction techniques lose precision.

First, we consider bulk modes in the thermodynamic limit, where boundaries are far from the region of interest and can therefore be neglected.
In Euclidean lattices, there should be $d$ branches of gapless (acoustic) phonon modes, where $d$ is the spatial dimension. These modes are the Goldstone excitations associated with the spontaneous breaking of $d$ translational symmetries by lattice formation and therefore remain gapless as required by the Goldstone theorem~\cite{goldstone1961field,nambu1960axial}. In contrast, we find that this paradigm breaks down in hyperbolic lattices due to the holonomy intrinsic to the vector nature of phonon fields. In two-dimensional hyperbolic lattices, although translational symmetries are likewise broken, the Goldstone theorem no longer ensures gapless modes. For coordination number five or higher, the bulk phonon spectrum becomes gapped, and the density of states (DOS) vanishes below a threshold frequency.
While zero-energy modes corresponding to uniform translations remain, they fail to form dispersive phonon bands with finite DOS, leaving a finite bulk gap—a fundamental departure from the behavior of Euclidean crystals. Representative examples such as the $\{3,7\}$ and $\{8,8\}$ lattices illustrate this bulk phonon gap.

In addition, we examine the role of boundaries by allowing phonon modes to reach the boundary. Remarkably, once boundary effects are included, a macroscopic number of states—scaling proportionally with the system size—emerges inside the bulk gap.
Unlike the subextensive in-gap edge modes of topological bands in flat space, where the ratio of edge to bulk modes vanishes in the thermodynamic limit, the finite boundary-to-bulk ratio of hyperbolic lattices alone is sufficient to dramatically reshape the spectrum, filling the bulk gap with an extensive number of boundary modes.
This behavior corresponds to an ``insulating bulk with a conducting boundary” at low frequencies—a hallmark of curvature-induced phononic organization. Even more strikingly, the zero-frequency modes themselves are already extensive: a finite fraction of all phonon modes reside on the boundary at $\omega = 0$, and this fraction remains finite even in the thermodynamic limit. This counting is strongly reminiscent of the particle occupation counting in Bose–Einstein condensation, where a finite fraction of particles occupy the ground state. 

\begin{figure*}[ht]
    \centering
    \includegraphics[width=\linewidth]{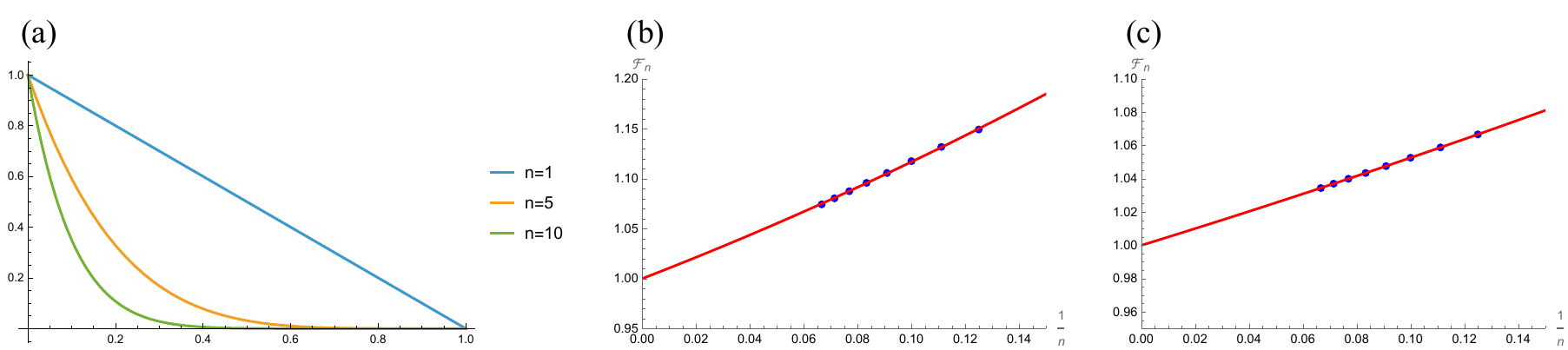}
    \caption{(a) The figure of $f_{n}(x)=(1-x)^{n}$ for $n=1,5,10$. As $n$ increase, $f_{n}(x)$ are more and more localized. (b) The triangular lattice is fitted with Eq.~\eqref{eq:fittingformula} by $\Delta=0$, $r=0.0411\pm0.0557$. The theoretical result is $\Delta=0$ and $r=0$. (c) The square lattice is fitted with Eq.~\eqref{eq:fittingformula} by $\Delta=0$, $r=-0.5018\pm 0.0018$. The theoretical result is $\Delta=0$ and $r=-0.5$.}
    \label{fig:EuclideanDOSFitting}
\end{figure*}

\noindent{\it The failure of the Goldstone theorem for vector fields in curved spaces}---Before presenting the theoretical analysis, we first provide a physical picture to understand why the Goldstone theorem fails in our study of vector fields on curved-space lattices.

We begin by distinguishing between bulk and edge modes. By analogy with flat space, bulk modes in curved lattices are defined as eigenmodes whose amplitudes remain bounded in all spatial directions, i.e., they are spatially extensive. In contrast, modes whose amplitudes grow or decay exponentially in certain directions are identified as edge or corner modes, whose weight is localized near parts of the boundary. From a group-theoretic perspective, bulk modes correspond to {\it unitary} representations of the translation group, whose unitarity ensures that translation preserves the mode amplitude, producing extended bulk states. Boundary modes, on the other hand, arise from {\it nonunitary} representations, for which translation induces exponential amplification or attenuation—thereby confining the mode to the boundary.

At a deeper level, the Goldstone theorem relies on two logically distinct ingredients: (1) the existence of a zero-energy mode associated with spontaneous symmetry breaking, and (2) the emergence of a gapless branch of excitations in the bulk. The first statement is general and insensitive to the geometry of space, whereas the second holds rigorously only under specific conditions, such as in flat space.

In both flat and curved lattices, spontaneous symmetry breaking inevitably generates zero-energy Goldstone modes corresponding to uniform, energy-neutral deformations of the order parameter. In flat space, these zero modes belong to unitary representations of the translation group and thus appear within the bulk spectrum—for instance, in phononic systems they constitute the $k=0$ limit of the acoustic branch, corresponding to uniform translations.

In curved space, spontaneous symmetry breaking likewise produces zero-energy deformations (e.g., uniform translations remain zero-energy phonon modes). However, because phonon displacements are vector fields that undergo nontrivial parallel transport, these Goldstone modes transform under a three-dimensional {\it nonunitary} irreducible representation of the translation group. As a result, they are topologically disconnected from the manifold of unitary representations that define the bulk bands. Since the bulk phonon spectrum is governed by unitary irreps, these nonunitary Goldstone modes are absent from the bulk bands. Consequently, while spontaneous symmetry breaking still guarantees the existence of zero-energy deformations, it no longer ensures a gapless excitation spectrum—the Goldstone theorem thus fails in curved spaces.

\noindent{\it Probing the bulk gap}---In this section, we present the technique we developed to probe whether a bulk gap exists near zero frequency. For a hyperbolic lattice, it is practically impossible to obtain all phonon branches, since the non-Abelian nature of the curved-space translation group gives rise to infinitely many higher-dimensional irreducible representations~\cite{Cheng_2022}. To overcome this difficulty, we adopt an alternative approach based on the energy moments of the spectrum.
We consider a lattice of $M$ particles of mass $m$, connected by Hookean springs with stiffness $k_e$. The corresponding dynamical matrix of the spring network is denoted by $D$, whose eigenvalues $E$ determine the phonon frequencies via $E = m\omega^2$. The $n$-th moment of the eigenenergy spectrum is defined as
\begin{equation}\label{eq:momentexpression_1}
\langle E^n\rangle = \int_{0}^{\infty} E^n \rho(E) dE
\end{equation}
where $\rho(E)$ is the DOS.
Because the moments $\langle E^n\rangle$ encode information about $\rho(E)$, they can be used to probe the existence of spectral gaps without explicitly computing all irreducible representations of the curved-space translation group.
Equivalently, the $n$-th moment can be computed directly from the dynamical matrix as
\begin{equation}\label{eq:momenttrace_2}
\langle E^n\rangle = \frac{1}{2M}\,\mathrm{Tr}\, D^n
= \frac{1}{M} \sum_{i=1}^{M}\frac{(D^n)_{i_{1},i_{1}} + (D^n)_{i_{2},i_{2}}}{2},
\end{equation}
where $i = 1, 2, \ldots, M$ labels the lattice sites, and the indices $1$ and $2$ denote two orthogonal translational degrees of freedom at each node. In the thermodynamic limit, focusing only on bulk sites, all sites become equivalent owing to lattice translational symmetry. Consequently, all terms in the sum on the right-hand side of Eq.~\eqref{eq:momenttrace_2} are identical, and the expression simplifies to
\begin{equation}\label{eq:momenttrace_3}
\langle E^n\rangle = \frac{(D^n)_{i_{1},i_{1}} + (D^n)_{i_{2},i_{2}}}{2},
\end{equation}
where $i$ can be any arbitrary bulk site.

Because the model only involves nearest-neighbor couplings,  $\rho(E)$ vanishes when $E > E_{0}$ for some $E_{0}$ sufficiently large (see Supplementary Materials for the choice of $E_{0}$), and the integration range in Eq.~\eqref{eq:momentexpression_1} can be truncated to $[0, E_0]$. Within this range, the function $(1 - E/E_0)^n$ is non-negative and develops an increasingly sharp peak at $E=0$ as $n \to \infty$
[Fig.~\ref{fig:EuclideanDOSFitting}(a)]. We thus define
\begin{equation}\label{eq:localizationformula}
    \mathcal{E}_n = \int_{0}^{E_{0}} (1 - E/E_{0})^n \rho(E)dE,
\end{equation}
where the weight function $(1 - E/E_{0})^n$ emphasizes the low-energy region of the spectrum and enables us to probe the behavior of the DOS near $E = 0$ for sufficiently large $n$.

If there exists a gap $\Delta$ near $E = 0$
\begin{equation}
    \rho(E) =
    \begin{cases}
    0, & \text{if } E< \Delta,\\
    \text{const}\times (E - \Delta)^{r}, & \text{if } \Delta<E\ll E_{0}.
    \end{cases}
\end{equation}
By the Weierstrass approximation theorem, $\rho(E)$ can be expanded as
\begin{equation}\label{eq:polynomialapproximationofrho}
    \rho(E)=\sum_{j=0}^{\infty}c_{j}\left(\frac{E/E_{0}-\Delta/E_{0}}{1-\Delta/E_{0}}\right)^{r+j}
\end{equation}
Defining $\mathcal{F}_{n} = \mathcal{E}_{n-1}/\mathcal{E}_n$ and expanding $\mathcal{F}_{n}$ in powers of $1/n$ around $n \to \infty$ yields
\begin{equation}\label{eq:fittingformula}
    \mathcal{F}_{n}=\frac{1}{1-\Delta/E_{0}}+\frac{1+r}{1-\Delta/E_{0}}\frac{1}{n}+\frac{c_{1}(1+r)}{c_{0}(1-\Delta/E_{0})}\frac{1}{n^{2}}+\mathcal{O}(\frac{1}{n^{3}}).
\end{equation}
The spectral gap $\Delta$ can then be extracted by fitting the data set $\{\mathcal{F}_{n}\}_{n=N_{0}}^{N_{1}}$, where $N_{0}$ and $N_{1}$ are chosen large enough that the higher-order correction terms in Eq.~\eqref{eq:fittingformula} become negligible. Ideally, once $N_{0}$ and $N_{1}$ are fixed, expanding $\mathcal{F}_{n}$ up to order $1/n^{N_{1}-N_{0}-1}$ gives the best estimate of $\Delta$ (see Supplementary Materials). In practice, however, the data ${\mathcal{F}_{n}}$ contain small numerical noise, and the optimal expansion order is determined empirically by the noise level (details in Supplementary Materials).

To validate the method, we benchmarked it against the phonon DOS of Euclidean triangular and square lattices in the low-frequency regime. For Bravais lattices in Euclidean space, the right-hand side of Eq.~\eqref{eq:momenttrace_3} can be obtained from the universal spectral moment theorem~\cite{Cheng_2024}. We computed the spectral moments up to $N_{1} = 15$ for both lattices. The low-frequency DOS scales as $\rho(E) \propto \text{constant}$ for the triangular lattice and $\rho(E) \propto E^{-1/2}$ for the square lattice.
By imposing the non-negativity condition $\Delta \geq 0$ through $\Delta/E_{0} = e^{-u}$ and taking $E_{0} = 6$ for the triangular lattice and $E_{0} = 2$ for the square lattice, we fit the data $\{\mathcal{F}_{n}\}_{n=8}^{15}$ using Eq.~\eqref{eq:fittingformula} [Figs.~\ref{fig:EuclideanDOSFitting}(b),(c)]. Under a $99\%$ confidence level, we obtain $\Delta = 0$, $r = 0.0411 \pm 0.0557$ for the triangular lattice, and $\Delta = 0$, $r = -0.5018 \pm 0.0018$ for the square lattice—both in excellent agreement with theory.

\begin{figure}[ht]
    \centering
    \includegraphics[width=\linewidth]{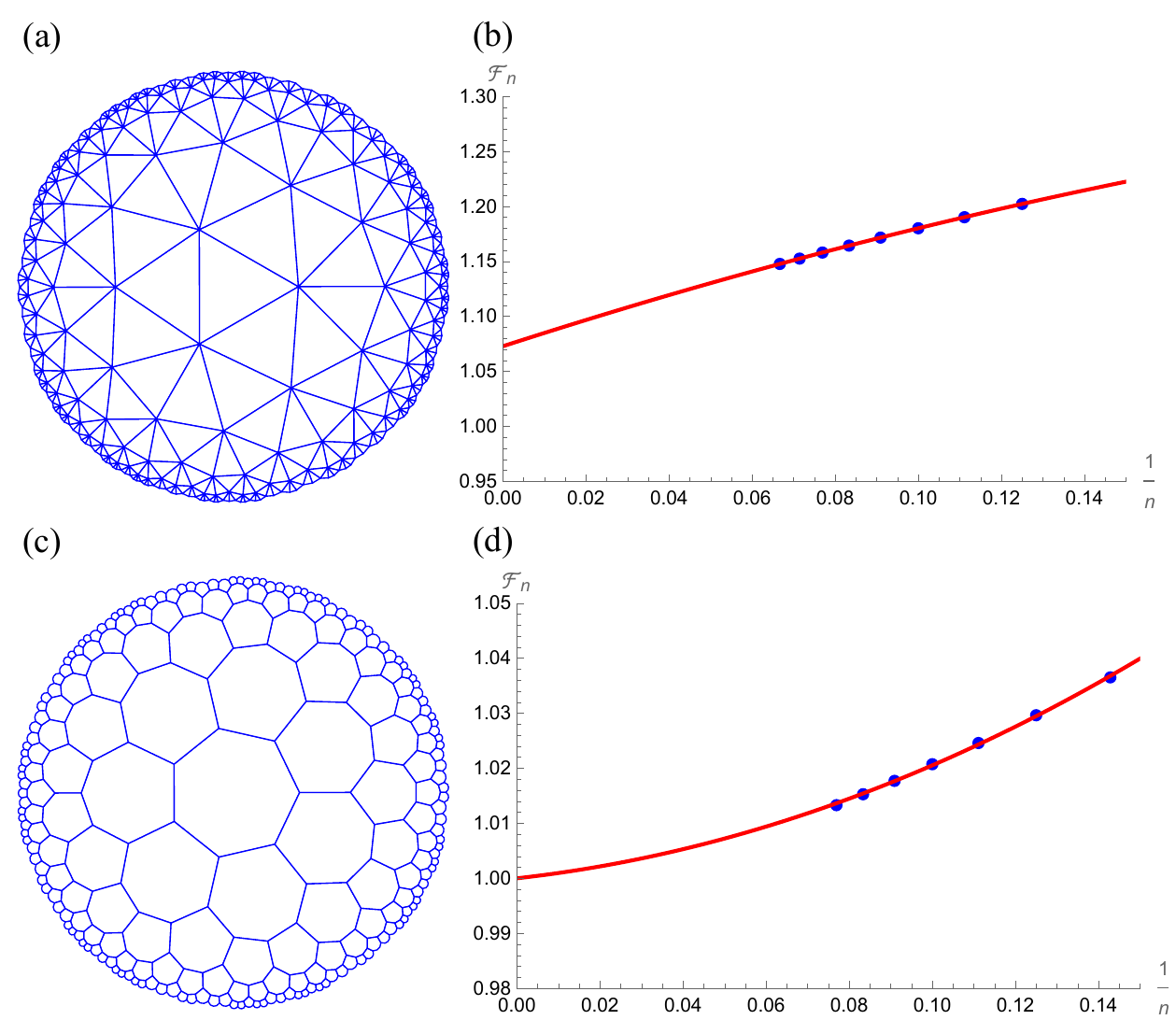}
    \caption{The hyperbolic $\{3,7\}$ and $\{7,3\}$ lattices and their low frequency phonon density of states property. (a) The elastic hyperbolic $\{7,3\}$ lattice in the Poincaré disk model. (b) Fitting $\{\mathcal{F}_{n}\}_{n=8}^{15}$ with respect to $1/n$ using Eq.~\eqref{eq:fittingformula} for the hyperbolic $\{3,7\}$ lattice. (c) The elastic hyperbolic $\{7,3\}$ lattice in the Poincaré disk model. (b) Fitting $\{\mathcal{F}_{n}\}_{n=7}^{13}$ with respect to $1/n$ using Eq.~\eqref{eq:fittingformula} for the hyperbolic $\{7,3\}$ lattice, $\omega_{g}=0$.}
    \label{fig:73and37phononDOS}
\end{figure}

\noindent{\it The insulating bulk in hyperbolic lattices}---Here we demonstrate the failure of the Goldstone theorem in the phonon spectrum of hyperbolic lattices. We begin with the $\{3,7\}$ tessellation [Fig.~\ref{fig:73and37phononDOS}(a)], where a point particle of mass $m$ is placed at each node, and neighboring nodes are connected by springs of stiffness $k_e$. Using the method described in the previous section, we numerically compute the first 15 moments of the phonon modes.
Specifically, we construct a finite $\{3,7\}$ lattice with eight layers [the first five shown in Fig.~\ref{fig:73and37phononDOS}(a)]. The finite dynamical matrix $D_f$ encodes the connectivity of this lattice: the diagonal element of $D_f^{n}$ corresponding to the $i^{\text{th}}$ ($i=j_{1},j_{2}$) degree of freedom at node $j$ counts the total weight of loops of length $n$ that start and end at node $j$. Choosing $j$ as the central node ensures that loops of length shorter than 15 do not reach the boundary. Consequently, for all $i=j_{1},j_{2}$, we have $[D^n]_{ii} = [D_f^n]_{ii}$, where $D$ is the dynamical matrix of the infinite lattice.

Setting $k_e = 1$ and $m = 1$, we fit the numerical data $\{\mathcal{F}_n\}_{n=8}^{15}$ using Eq.~\eqref{eq:fittingformula}, with an optimal expansion order of 2. The fit (with $99\%$ confidence level) yields an energy gap $\Delta/E_0 = 0.0687 \pm 0.0007$, where $E_0 = 9.1984 k_e$. This corresponds to a phonon frequency gap
\begin{equation}\label{eq:frequencygap}
\omega_g = (0.795 \pm 0.004)\sqrt{k_e/m},
\end{equation}
indicating an insulating bulk at zero frequency.

For generic hyperbolic lattices, the number of sites grows much faster with layer number than in the $\{3,7\}$ case, making the real-space method computationally expensive. A more efficient approach for evaluating spectral moments is introduced in the Supplementary Materials and applied to the $\{8,8\}$ lattice. There, we confirm that the phonon spectrum of the infinite $\{8,8\}$ lattice also exhibits a finite spectral gap at $\omega = 0$.

The origin of this gap lies in two essential ingredients: negative curvature and a lattice coordination number $z$ exceeding $2d$.
The necessity of negative curvature can be understood by contradiction. If the same network connectivity were embedded in the Euclidean plane, long-wavelength eigenmodes continuously connected to the Goldstone modes would appear, leading to a finite DOS arbitrarily close to $\omega = 0$. This argument breaks down in hyperbolic space. Because of the negative curvature, the Goldstone modes of a hyperbolic lattice belong to nonunitary representations of the translation group, with amplitudes that grow exponentially along certain directions. Physical phonon eigenmodes—bounded throughout hyperbolic space—therefore cannot be continuously connected to these zero-frequency Goldstone modes.

The necessity of $z > 2d$ follows from comparison with the underconstrained $\{7,3\}$ lattice [Fig.~\ref{fig:73and37phononDOS}(c)], which is the graph dual of the $\{3,7\}$ tiling. In the $\{7,3\}$ lattice, each mass point is connected to only three springs. According to Maxwell’s counting argument, each spring provides one constraint, while each site has $d = 2$ translational degrees of freedom. Thus, with $z = 3$, each particle experiences on average only $z/2 = 3/2$ constraints—fewer than its degrees of freedom—rendering the lattice underconstrained and mechanically floppy. Due to the presence of floppy modes, an underconstrained lattice naturally supports zero-frequency and low-frequency excitations. Applying the same analysis as above, we find that the $\{7,3\}$ lattice exhibits no spectral gap in its phonon DOS at $\omega = 0$. This result reinforces that both negative curvature and overconstraint ($z > 2d$) are necessary conditions for the emergence of an insulating bulk in hyperbolic lattices.

\noindent{\it The conducting edge}---Given the extensive boundary characteristic of hyperbolic lattices, a natural question arises regarding whether the spectral gap at $\omega=0$, observed in infinite systems, persists in finite patches of the lattice. Here, we show that the gap does not survive finite-size effects: by a general counting argument applied to non-triangular hyperbolic lattices, coupled with numerically diagonalizing the dynamical matrix of the finite $\{3,7\}$ lattice, we demonstrate that extensive boundaries necessarily close the gap.

\begin{figure*}[ht]
    \centering
    \includegraphics[width=1.0\linewidth]{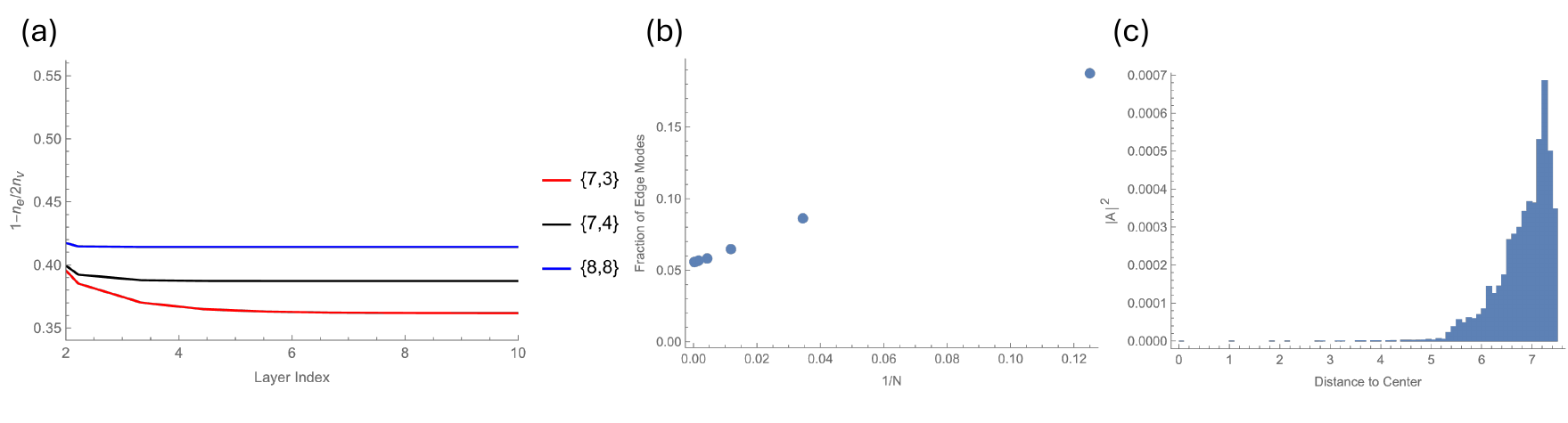}
    \caption{(a) The lower bound on the fraction of floppy modes for hyperbolic $\{7,3\}$, $\{7,4\}$ and $\{8,8\}$ lattices. (b) Fraction of phonon eigenmodes below the band gap as a function of $1/N$ in the $\{3, 7\}$ lattice, where $N$ is the number of nodes in the network. (c) The shape of a random eigenmode of the $\{3, 7\}$ lattice in the phonon band gap at $\omega=0$. The figure is the plot of the norm-squared of the amplitude verses distance averaged over bins of size $0.2 R$, where $R$ is the radius of curvature of the hyperbolic space.}
    \label{fig:edgemodes}
\end{figure*}

We begin by introducing a counting argument. Let $n_{e}$ the the number of edges in a finite patch of hyperbolic $\{p,q\}$ lattice, and let $n_{v}$ be the number of nodes in the same lattice. The portion of zero modes is bounded from below by $1-n_{e}/2n_{v}$. For non-triangular hyperbolic lattices, this value is non-zero even when the number of layers of the lattice tends to infinity (Fig.~\ref{fig:edgemodes} (a), see Supplemental Materials for a recursive method computing $n_{e}$ and $n_{v}$). If the bulk phonon spectrum exhibits a gap at $\omega=0$ (as the $\{8,8\}$ lattice), then all the non-trivial zero modes must be localized at the boundary. 

This ratio, however, necessarily approaches zero for triangular hyperbolic lattices as these lattices are rigid and do not support non-trivial zero modes. Thus, we study the finite $\{3,7\}$ lattice through numerical diagonalization of the dynamical matrix. We find that the fraction of modes $f$ below this estimated gap approaches $5-6\%$ of the system size as system size increases (see Fig.~\ref{fig:edgemodes} (b)). Lying in the bulk band gap, these modes are necessarily edge modes (Fig.~\ref{fig:edgemodes} (c)).

\noindent{\it Conclusion}---In this Letter, we investigate the interplay between negative curvature and extensive boundary on phononic properties of hyperbolic lattices. Through moment analysis of the density of states, we demonstrate that, in sharp contrast to Bravais lattices in the Euclidean space, the bulk phonon spectrum of infinite over-constrained hyperbolic lattices features a finite spectral gap at zero frequency, a direct consequence of the negative curvature inherent to hyperbolic space and the over-constraint-ness of some hyperbolic lattices. We further show that any finite hyperbolic lattice necessarily closes this gap due to the presence of an extensive boundary, giving rise to an abundance of low-frequency edge modes. These findings open new avenues for exploring phononic topological states in hyperbolic lattices.

It is also worth mentioning that the right-hand side of Eq.~\eqref{eq:momenttrace_3} depends exclusively on closed loops of length $n$, meaning our method for characterizing the low-frequency phonon density of states relies solely on the local properties of the mechanical network. Consequently, even though low-frequency phonons are typically long-wavelength modes, their density of states can still be determined from local mechanical information in systems with long-range order. This fact implies that the applicability of this method is not limited to hyperbolic lattices but extends to other homogeneous systems with only local fluctuations, such as fractals and locally disordered materials, such as jammed amorphous solids and hyperuniform disordered solids.

\newpage
\begin{widetext}
\section{Supplementary Materials}

\subsection{\label{si:73_goldstonemodes} Goldstone Modes Carry $3$D Non-unitary Representations}

Here, we explain the necessity of 3D nonunitary representations in giving rise to goldstone modes. $\text{SO}(1, 2)$ acts on the hyperbolic plane embedded in Minkowski space. Its action correspond to orientation-preserving isometries of the plane. $\mathrm{SO}(1, 2)$ has 3 infinitesimal generators, which we label $K_1, K_2, K_3$; let $K = (K_1, K_2, K_3)$. A goldstone mode $v_\alpha$ associated to the generator $\alpha \cdot K$, where $\alpha = (\alpha_1, \alpha_2, \alpha_3)$ is a unit-norm vector, is given by
\begin{equation}
v_\alpha(r) = \lim_{t \to 0} \frac{e^{t \alpha \cdot K} r - r}{t} = (\alpha \cdot K) r
\end{equation}
for any lattice site $r$. There are $3$ linearly independent $K_i$, so there are $3$ linearly independent goldstone modes associated to the $3$ generators of $\mathrm{SO}(1, 2)$: $v_i = v_{K_i}$ for $i = 1, 2, 3$. Now, let $g = e^{\beta \cdot K} \equiv e^X$ be a hyperbolic translation in $\mathrm{SO}(1, 2)$. Consider the displacement field $g v_i$ defined by
\begin{equation}
    (g v_i)(r) \equiv g (v_i(g^{-1}r))
\end{equation}
Its displacement at site $g r$ is
\begin{equation}
(g v_i)(gr)=g(v_{i}(r))= g K_i r = g K_i g^{-1} g r.
\end{equation}
We will write $g K_i g^{-1}$ as a linear combination of $K_1, K_2, K_3$. Direct calculation shows that
\begin{equation}
[K_i, K_j] = \sum_k(K_i)_{k j} K_k.
\end{equation}
Thus for any element $Y \in \mathfrak{so}(1, 2)$ of the Lie algebra of $\operatorname{SO}(1, 2)$, we have
\begin{equation}
[Y, K_j] = Y_{k j} K_k.
\end{equation}
Applying the Baker-Hausdorff formula, we see that
\begin{equation}
g K_i g^{-1} = e^X K_i e^{-X} = \left(e^X \right)_{j k} K_j = g_{j k} K_j.
\end{equation}
Hence, denoting the $(i, j)$-entry of $g$ by $g_{j i}$, we have
\begin{equation}
(g v_i)(gr) = \sum_j g_{j i} v_j(gr).
\end{equation}
Since $g$ and $r$ are arbitrary, we have
\begin{equation}
    g v_i = \sum_j g_{j i} v_j.
\end{equation}
It follows that the Goldstone mode carries the tautological representation of hyperbolic translations, i.e. the representation $\rho$ that maps an orientation preserving isometry of the hyperbolic plane to its matrix in the hyperboloid model. Since $\rho$ is a non-unitary $3$D irreducible representation,  we conclude that Goldstone modes are described by high dimensional non-unitary irreducible representations.

\subsection{\label{si:estimateEbound} An Estimate for $E_{0}$}

In the main text we mentioned that the density of states $\rho(E)$ for any $\{p,q\}$ lattice with only nearest neighbor coupling has the property $\sup\{E\mid \rho(E)>0\}<+\infty$. Define $E_{1}$ to be
\begin{equation}
E_{1}=\inf\Biggl\{\sqrt{\left(\sum_{k_{s}} \left(D_{j_{1}k_{s}}^{2}+D_{j_{2}k_{s}}^{2}\right)\right)(q+1)}\Biggl\},
\end{equation}
where $k_{s}$ is the $s=1,2$ orthogonal degrees of freedom at node $k$, $j$ is the node at the origin and the infimum is taken over all choices of local orthogonal degrees of freedom. We show that when $E>E_{1}$, $\rho(E)=0$.

To prove this statement, let $\phi$ be an eigenstate of $D$ with eigenvalue $\epsilon$, i.e. $D \phi = \epsilon \phi$. We need to show that $\epsilon \leq E_{1}$. Let $\phi_{j_{s}}$ be the component of $\phi$ associated to the $s=1,2$ orthogonal degrees of freedom of node $j$. Since we are focusing on the phonon modes whose amplitude at any given node is bounded and the two degrees of freedom at any node are orthogonal, $A_{j}\equiv\phi_{j_{1}}^{2}+\phi_{j_{2}}^{2}$ is also bounded and independent of the choice of local degree of freedom. For any $\delta>0$, we could always choose a lattice site $j$ such that 
\begin{equation}
    A_{j}\geq \sup\{A_{r}|\; r \text{ is a lattice site}\}-\delta.
\end{equation}
By translational symmetry, we could always assume this node to be the node at the origin.
From $D\phi=\epsilon\phi$, we have
\begin{equation}
\begin{aligned}
&\epsilon \phi_{j_{1}} = \sum_{k_{s}} D_{j_{1}k_{s}} \phi_{k_{s}}\\
&\epsilon \phi_{j_{2}} = \sum_{k_{s}} D_{j_{2}k_{s}} \phi_{k_{s}}
\end{aligned}
\end{equation}
Letting $B=\{k\mid\exists\, s\in [1,2] \text{ s.t. } D_{j_{1}k_{s}}\ne 0 \text{ or } D_{j_{2}k_{s}}\ne 0\}$, we get
\begin{equation}
\begin{aligned}
\epsilon^{2} (\phi_{j_{1}}^{2}+\phi_{j_{2}}^{2}) &= \left(\sum_{k_{s}} D_{j_{1}k_{s}}\phi_{k_{s}}\right)^{2}+\left(\sum_{k_{s}} D_{j_{2}k_{s}}\phi_{k_{s}}\right)^{2}\\
&\leq \left(\sum_{k\in B}\sum_{s=1}^{2} D_{j_{1}k_{s}}^{2}\right)\sum_{k\in B}\left(\phi_{k_{1}}^{2}+\phi_{k_{2}}^{2}\right)+\left(\sum_{k\in B}\sum_{s=1}^{2} D_{j_{2}k_{s}}^{2}\right)\sum_{k\in B}\left(\phi_{k_{1}}^{2}+\phi_{k_{2}}^{2}\right)\\
&\leq \left(\sum_{k\in B}\sum_{s=1}^{2} \left(D_{j_{1}k_{s}}^{2}+D_{j_{2}k_{s}}^{2}\right)\right)(A_{j}+\delta)|B|,
\end{aligned}
\end{equation}
where $|B|\leq q+1$ is the number of elements in the set $B$. Since $\delta>0$ is arbitrary, we have
\begin{equation}\label{eq:SIupperbound}
\begin{aligned}
\epsilon &\leq \sqrt{\left(\sum_{k\in B}\sum_{s=1}^{2} \left(D_{j_{1}k_{s}}^{2}+D_{j_{2}k_{s}}^{2}\right)\right)|B|}\\
&\leq \sqrt{\left(\sum_{k\in B}\sum_{s=1}^{2} \left(D_{j_{1}k_{s}}^{2}+D_{j_{2}k_{s}}^{2}\right)\right)(q+1)}\\
&= \sqrt{\left(\sum_{k_{s}} \left(D_{j_{1}k_{s}}^{2}+D_{j_{2}k_{s}}^{2}\right)\right)(q+1)},
\end{aligned}
\end{equation} 
Since the inequality Eq.~\eqref{eq:SIupperbound} is true for all node $j$ and all choices of local orthogonal degrees of freedoms, we have
\begin{equation}
    \epsilon\leq\inf\Biggl\{\sqrt{\left(\sum_{k_{s}} \left(D_{j_{1}k_{s}}^{2}+D_{j_{2}k_{s}}^{2}\right)\right)(q+1)}\Biggl\}= E_{1}
\end{equation}
In practice, we can fix arbitrary orthogonal local degrees of freedom and choose the $E_{0}$ mentioned in the main text to be 
\begin{equation}
    E_{0}=\sqrt{\left(\sum_{k_{s}} \left(D_{j_{1}k_{s}}^{2}+D_{j_{2}k_{s}}^{2}\right)\right)(q+1)}\geq E_{1}.
\end{equation}

\subsection{Estimating $\mathcal{F}_{\infty}$ from Samples $\{\mathcal{F}_{n}\}_{n=N_{1}}^{N_{2}}$: Analytic Framework, Weighting, and Model Selection}

In the application considered in the main text, the data is encoded by a sequence of ``moments'' $\{\langle E^{m}\rangle\}_{m\ge0}$ and a known constant $E_0>0$, and we work with
\begin{equation}
\mathcal{E}_n \;=\; \sum_{m=0}^{n} \binom{n}{m}\,\frac{\langle E^{m}\rangle}{(-E_0)^m},\qquad
\mathcal{F}_n \;=\; \frac{\mathcal{E}_{n-1}}{\mathcal{E}_n}\quad(n\ge2),
\label{eq:defTR}
\end{equation}
which are then viewed as samples $\bigl(1/n,\mathcal{F}_n\bigr)$ over an index set $\mathcal{I}$ (e.g., $\mathcal{I}=\{N_{0}, N_{1}+1,\dots,N_{1}\}$). Let $f(1/n)=\mathcal{F}_{n}$. The fitting problem can be reformulated as seeking $f(0)$ given values of the function $f$ at $\{x_{n}=1/n \mid N_0 \leq n \leq N_1\}$.
Assume $f$ extends analytically to a complex neighborhood of $x=0$, namely the constructed map $x\mapsto f(x)$ is analytic at $0$. $f$ can be parameterized by the low-order Taylor expansion
\begin{equation}
    f(x)\;=\;\frac{1}{1-\Delta/E_{0}}\;+\;\frac{1+r}{1-\Delta/E_{0}}\,x
    \;+\;\sum_{i=2}^{k} a_i x^i,
\label{eq:model}
\end{equation}
where $a_2,\dots,a_k$ encode higher-order analytic corrections, and $\Delta$ and $r$ are defined by 
\begin{equation}
    \rho(E) =
    \begin{cases}
    0, & \text{if } E< \Delta,\\
    \text{const}\times (E - \Delta)^{r}, & \text{if } \Delta<E\ll E_{0}.
    \end{cases}
\end{equation}
$f(0)$ is then obtained from the fitted parameters via Eq.~\eqref{eq:model}.

Treat $\langle E^{m}\rangle$ as independent with standard deviations $\sigma_m$ reflecting their quoted precision. Linear propagation from \eqref{eq:defTR} gives
\begin{align}
    \mathrm{Var}(\mathcal{E}_n) &= \sum_{m=0}^{n} c_{n,m}^2\,\sigma_m^2,\quad
    c_{n,m}\equiv \binom{n}{m}(-E_0)^{-m}, \label{eq:varT}\\
    \mathrm{Cov}(\mathcal{E}_{n-1},\mathcal{E}_n) &= \sum_{m=0}^{n-1} c_{n-1,m}\,c_{n,m}\,\sigma_m^2, \\
    \mathrm{Var}(\mathcal{F}_n) &\approx \frac{\mathrm{Var}(\mathcal{E}_{n-1})}{\mathcal{E}_n^2}
    +\frac{\mathcal{F}_n^2\,\mathrm{Var}(\mathcal{E}_n)}{\mathcal{E}_n^2}
    -\frac{2\mathcal{F}_n\,\mathrm{Cov}(\mathcal{E}_{n-1},\mathcal{E}_n)}{\mathcal{E}_n^2}. \label{eq:varR}
\end{align}
Weighted nonlinear least squares then minimizes
\begin{equation}
\min_{\Delta,m,\{a_i\}} \;\sum_{n\in\mathcal{I}} w_n\,
\bigl(\mathcal{F}_n-f(x_n)\bigr)^2,\qquad
w_n\equiv \mathrm{Var}(\mathcal{F}_n)^{-1}.
\label{eq:wnlls}
\end{equation}
This inverse-variance weighting is statistically optimal under independent Gaussian errors with heterogeneous variances and remains appropriate here due to the small propagated errors.

Because all $x_n>0$, estimating $f(0)$ is a left-extrapolation problem: higher $k$ reduces bias but inflates variance. We choose $k$ by minimizing a data-driven estimate of predictive risk using leave-one-out cross-validation (LOOCV). For each candidate $k$ we fit \eqref{eq:wnlls} on all points except $k$ to obtain $\widehat{\mathcal{F}}_{k,n}$, then define the (weighted) LOOCV score
\begin{equation}
\mathrm{LOOCV}(k)\;=\;\frac{1}{|\mathcal{I}|}\sum_{n\in\mathcal{I}}
\frac{\bigl(\mathcal{F}_n-\widehat{\mathcal{F}}_{k,n}(x_n)\bigr)^2}{\mathrm{Var}(\mathcal{F}_n)}.
\label{eq:loocv}
\end{equation}
We take $k^\star=\text{argmin}_k \mathrm{LOOCV}(k)$, with the constraint that the number of fitted coefficients ($2+(k-1)$) remains well below $|\mathcal{I}|$ to preserve identifiability and conditioning. This criterion balances the analytic bias–variance tradeoff inherent in extrapolating to $x=0$ and is asymptotically unbiased for the out-of-sample risk.

It is worth mentioning: (i) Under analyticity, Cauchy bounds imply that the truncation error in \eqref{eq:model} decays geometrically with $k$ until the variance term dominates; LOOCV detects the turnover without requiring the (unknown) radius of convergence. (ii) The weighting \eqref{eq:wnlls} calibrates residuals to the uncertainty scale set by \eqref{eq:varR}, ensuring that parameter estimates—and thus $f(0)$—are driven by the most informative indices. (iii) With $|\mathcal{I}|$ in the single digits, the optimal $k^\star$ is typically moderate (e.g., $k=2$), which we also observe empirically for the dataset used in the main text.

\subsection{Moment Computation in the $\{8, 8\}$ Lattice}
We describe the method we used for computing the moments of the $\{8, 8\}$ tiling in this section. For the infinite $\{8,8\}$ lattice, the dynamical matrix is~\cite{Cheng_2022} 
\begin{equation}
	D=D_0\otimes\sum_{t\in T}\ket{t}\bra{t}+\sum_{j=1}^{4}\left(D^R_j\otimes\sum_{t\in T}\ket{t}\bra{t\gamma_j}\right)+\sum_{j=1}^{4}\left(D^L_j\otimes\sum_{t\in T}\ket{t\gamma_j}\bra{t}\right), \label{SI:EQ:Dmatirx}
\end{equation}
where $\{\gamma_{i}\}_{i=1}^{4}$ are the generators of the translation group of the $\{8,8\}$ lattice (Ref.~\cite{Maciejko_2021} gives an explicit formula for $\gamma_{i}$). $D^{n}$ takes the general form
\begin{equation}
    D^{n}=D(n)_{0}\otimes\sum_{t\in T}\ket{t}\bra{t}+D_{\text{off}}(n),
\end{equation}
where $D_{\text{off}}(n)$ are the terms whose hopping between sites take the form $\sum_{t\in T}\ket{t}\bra{tg}$ or $\sum_{t\in T}\ket{tg}\bra{t}$ for some $g\ne\text{id}$. We have
\begin{equation}
    \langle E^{n}\rangle=\tr(D(n)_{0}).
\end{equation}
Now, the computation of $\langle E^{n}\rangle$ reduces to the enumeration of weighted loops (weighted by matrices $D_{0}, D_{j}^{R}$ and $D_{j}^{L}$). Such enumeration can be carried out efficiently.

\subsection{\label{si:recursive} Recursive Counting Formula}

In this section, we derive a recursive formula for counting the number of vertices $s \equiv n_v$ and edges $b \equiv  n_e$ as a function of system size. We define a layering of a graph $G$ as a sequence of complete subgraphs $G_n$ such that $G_{n - 1}$ is a subgraph of $G_n$, and $G_n \to G$ as $n \to \infty$. For any graph $G$, we let $V(G)$ denote the set of vertices and $E(G)$ the set of edges.

First, fix $\{p, q\}$ and assume $\frac{1}{p} + \frac{1}{q} \leq \frac{1}{2}$; that is, we do not consider spherical lattices, only hyperbolic and Euclidean. Let $G$ be the $\{p, q\}$ lattice. Let the zeroth layer, i.e. $G_0$, consist of a single vertex, say at the origin. $G_{n + 1}$ is determined by $G_n$ as follows:

\begin{itemize}
    \item[(1)] Let $V'_{n + 1}$ be the set of vertices $v \in V(G) \setminus V(G_n)$ such that there exists an edge $(v, v')$ for some vertex $v \in V(G_n)$.
    \item[(2)] Let $V''_{n + 1}$ be the set of vertices $v \in V(G) \setminus(V(G_n) \cup V_{n + 1}')$ such that there exists a cycle of $p$ vertices containing $v$ and some $v' \in V(G_n)$.
    \item[(3)] Set $V(G_{n + 1}) = V(G_n) \cup V'_{n + 1} \cup V''_{n + 1}.$
    \item[(4)] $G_{n + 1}$ is the complete subgraph of $G$ whose vertex set is $V(G_{n+1})$.
\end{itemize}

Now, we derive the counting formula for $p, q > 3$. Let $s_1(n) = |V_{n}'|$ and $s_2(n) = |V_{n}''|$; clearly, $s_1(1) = q$ and $s_2(1) =  (p - 3) q$. Let $s(n) = s_1(n) + s_2(n)$. $s_1(n)$ determines the number of vertices on the boundary of $G_n$ that are connected to $G_{n - 1}$ by a single edge, while $s_2(n)$ are those vertices of the boundary that complete all cycles of length $q$ along the boundary. Then, the total number of vertices is the sum of vertices along each boundary:
\[|V(G_n)| = \sum_{k = 0}^n s(k).\]
Similarly, let $b(n) = |E (G_n) \setminus E(G_{n - 1})|$ for $n > 1$ with $b(1) = |E(G_1)|$. $b(n + 1)$ is the number of edges added when going from the $n$th step to the $(n + 1)$-th step Thus,
\[|E(G_n)| = \sum_{k = 1}^n b(n).\]
It is straightforward to see that $b(1) =  (p - 1) q$. Now, for each vertex $v$ in $V_n'$, there are $(q - 3)$ vertices $v'$ in $V_{n + 1}'$ such that $\{v, v'\} \in E(G)$ when $p \geq 3$. Similarly, for each vertex $v$ in $V_n''$, there are $(q - 2)$ vertices $v'$ in $V_{n + 1}'$ such that $\{v, v'\} \in E$. By construction, these are all the vertices in $V_{n + 1}'$, so when $p \geq 3$,
\[s_1(n + 1) = (q - 3) s_1(n) + (q - 2) s_2(n).\]

For vertices in $V_{n + 1}''$ there are two types of cycles to consider: (1) those that do not share an edge with $G_n$, and (2) those that do share an edge with $G_n$. (1) are those $p$-gons that meet $G_n$ at only one vertex, and hence there are $(p - 3)$ additional vertices to add for each such $p$-gon and $s_1(n + 1) - s(n)$ such $p$-gons. (2) are those $p$-gons that meet $G_n$ at an edge. Because $q > 3$, there are $(p - 4)$ additional vertices to add for each $p$-gon, and there are $s(n)$ $p$-gons. Therefore,
\[s_2(n + 1)  = (p - 3) [s_1(n + 1) - s(n)] + (p - 4) s(n)\]

Next, we count the edges. There are edges connecting all the vertices in $V(G_{n + 1}) \setminus V(G_n)$ and there are edges for each vertex in $V_{n + 1}'$, so
\[b(n + 1) = s_1(n + 1) + s(n + 1) = 2 s_1(n + 1) + s_2(n + 1).\]
Clearly, $b(1) = 2 s_1(1) + s_2(1)$, so $b(n) = 2 s_1(n) + s_2(n)$ for all positive integers $n$. Together, $s_1, s_2, b$ give an exact counting of the vertices and edges.

When $p = 3$, the counting becomes much simpler. Again, let $s(n)$ denote the number of vertices on the boundary, i.e. $s(n) = |V(G_{n + 1}) \setminus V(G_n)|$ with $s(0) = 1$ and $s(1) = q$. It is also easy to see that $s(2) = q(q - 4)$, so we consider $n > 2$. There are two types of vertices $v \in G_n$ on the $n$th boundary: (1) there are three edges in $G_n$ containing $v$, and (2) there are four edges in $G_n$ containing $v$. The number of vertices of case (2) is $s(n - 1)$, since these are vertices that have a $3$-cycle sharing an edge with $G_{n - 1}$. The number of vertices of case (1) is obviously $s(n) - s(n - 1)$. $q > 5$ for hyperbolic/Euclidean tilings, so we add $q - 4$ vertices for each case (1) and $q - 5$ vertices for each case (2). Therefore,
\[\begin{split}
    s(n + 1) &= (q - 4)( s(n) - s(n -1)) + (q - 5) s(n - 1).
    \\ &=(q - 4) s(n) - s(n - 1).
\end{split}\]
The number of edges $b(n + 1)$ in $E(G_{n + 1}) \setminus E(G_n)$ is then sum of the following: the number of edges $s(n + 1)$ from one vertex in $V(G_{n})$ to one vertex $V(G_{n + 1})$; the number of edges $s(n + 1)$ connecting the vertices in $V(G_{n + 1})$ to other vertices in $V(G_{n + 1})$; and $s(n)$ edges for the vertices in $V(G_{n + 1})$ with $3$-cycles sharing an edge with $G_n$. Then,
\[b(n + 1) = 2 s(n + 1) + s(n).\]

\subsection{\label{si:fraction_rigid} Non-zero Fraction of Modes Below the Gap}

In this section, we provide justification for why the fraction of modes $f$ below the gap does not approach zero in the rigid $\{3, 7\}$ lattice. Let $f(N)$ be the fraction as a function of system size $N$. From Fig. \ref{fig:fraction_log}, we expect that $f(N) \log(N) \to \infty$ as $N \to \infty$. If $f(N) \to 0$ as $N \to \infty$, then we can estimate $f$ by a power law $f(N) = N^{-\alpha}$ for $\alpha > 0$. But then, $f(N) \log(N) = N^{-\alpha} \log(N) \to 0$ as $N \to \infty$ since $\alpha > 0$.

\begin{figure}[ht]
    \centering
    \includegraphics[width=0.4\linewidth]{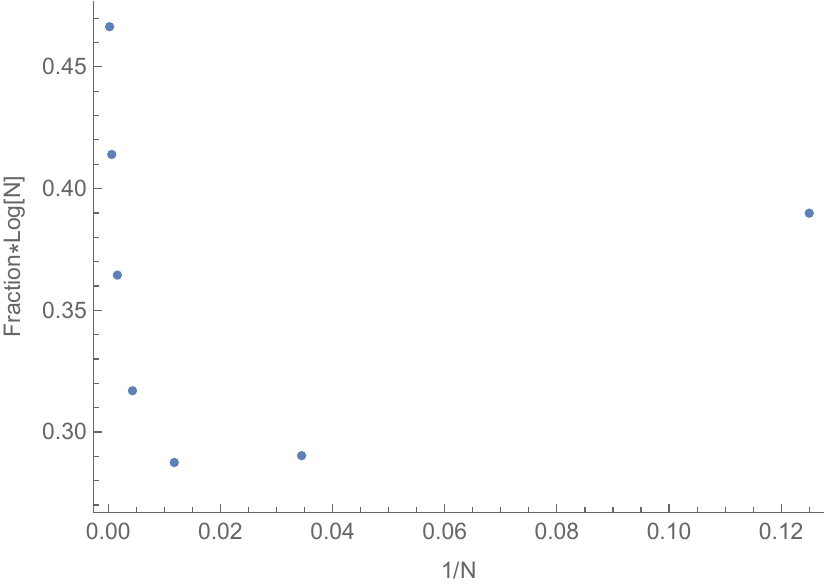}
    \caption{Plot of $f \log(N)$ as a function of $1/N$.}
    \label{fig:fraction_log}
\end{figure}

\subsection{\label{si:proofs} Non-rigidity of Finite Graphs}

In this section, we prove the following theorem:
\begin{theorem}
    Let $G = (V, E)$ be a finite planar graph. Suppose one face of $G$ consists of $\geq 3$ edges and every other face consists of $\geq 4$ edges. Then, $G$ is not $2$-rigid.
\end{theorem}

The following lemma is a well-known fact about rigidity:

\begin{lemma}
    \label{lemma:rigidity_counting}
    Let $G = (V, E)$ be a finite planar graph. If $2 |V| - |E| > 3$, then $G$ is not $2$-rigid.
\end{lemma}
\begin{proof}
    The compatibility matrix defines a map
    \[C : \RR^{|V|} \times \RR^{|V|} \to \RR^{|E|}\]
    whose kernel is the set of zero modes. By rank-nullity,
    \[\dim{\ker{C}} = 2 |V| - \dim{\im{C}} \geq 2 |V| - |E|.\]
    The rigid modes form a $3$-dimensional subspace of the kernel. Therefore, if $2 |V| - |E| > 3$, then $\dim{\ker{C}} > 3$, and hence there is at least one non-rigid zero mode.
\end{proof}

We can now prove Theorem S1:
\begin{proof}[Proof of Theorem S1]
    Let $F$ be the set of faces, and let $N_e(f)$ denote the number of edges associated to a face $f \in F$. Each edge is a member of exactly two faces, so
    \[\sum_{f \in F} N_e(f) = 2 |E|.\]
    Let $\tilde{f} \in F$ with $N_e(\tilde{f}) \geq 3$ be the triangular face if it exists in $G$. Then, for all $f \in F \setminus \{\tilde{f}\}$, $N_e(f) \geq 4$, and hence
    \[|E| = \frac{1}{2} \sum_{f \in F} N_e(f) \geq \frac{1}{2} \left( 4(|F|  - 1) + 3 \right) = 2 |F| - \frac{1}{2}.\]
    The Euler characteristic of $G$ is
    \[|F| - |E| + |V| = 2.\]
    Therefore,
    \[\begin{split}
        2 |V| - |E| &= |E| - 2 |F| + 4 
        \\ &\geq \left(2 |F| - \frac{1}{2} \right) - 2|F| + 4 
        \\ &\geq 3 + \frac{1}{2} > 3.
    \end{split}\]
    Finally, apply Lemma \ref{lemma:rigidity_counting}.
\end{proof}

\end{widetext}
\bibliography{apssamp}
\end{document}